TITLE: CARCERAL ALGORITHMS AND THE HISTORY OF CONTROL: AN ANALYSIS OF THE PENNSYLVANIA ADDITIVE CLASSIFICATION TOOL

**Authors and affiliations:**

Vanessa Massaro, Bucknell University
Swarup Dhar, Bucknell University
Darakhshan Mir, Bucknell University
Nathan C. Ryan, Bucknell University

**Abstract**
Scholars have focused on algorithms used during sentencing, bail, and parole, but little work explores what we call carceral algorithms that are used *during* incarceration. This paper is focused on the Pennsylvania Additive Classification Tool (PACT) used to classify prisoners' custody levels while they are incarcerated. Algorithms that are used during incarceration warrant deeper attention by scholars because they have the power to enact the lived reality of the prisoner. The algorithm in this case determines the likelihood a person would endure additional disciplinary actions, can complete required programming, and gain experiences that, among other things, are distilled into variables feeding into the parole algorithm. Given such power, examining algorithms used on people currently incarcerated offers a unique analytic view to think about the dialectic relationship between data and algorithms. Our examination of the PACT is two-fold and complementary. First, our qualitative overview of the historical context surrounding PACT reveals that it is designed to prioritize incapacitation and control over rehabilitation. While it closely informs prisoner rehabilitation plans and parole considerations, it is rooted in population management for prison securitization. Second, on analyzing data for 146,793 incarcerated people in PA, along with associated metadata related to the PACT, we find it is replete with racial bias as well as errors, omissions, and inaccuracies. Our findings to date further caution against data-driven criminal justice reforms that rely on pre-existing data infrastructures and expansive, uncritical, data-collection routines.

**Introduction**

      This paper focuses on an algorithm that is used to classify prisoners' custody levels while they are incarcerated: the Pennsylvania Additive Classification Tool (PACT). Scholars have focused on algorithms used during sentencing, bail, and parole, but little work explores the algorithms that are used *during* the period of incarceration. This carceral algorithm was implemented thirty years ago, in 1991, to generate a raw score that determines a person's custody level. Custody level dictates what type of prison a person is assigned to (in Pennsylvania this ranges from minimum to maximum security housing of prisoners). The PACT is an algorithm that has the power to enact the lived reality of the prisoner and thus determines the likelihood a person would endure additional disciplinary actions, can complete required programming and other experiences that are then distilled into variables feeding into the parole algorithm (see Figure 4). Given such power, examining algorithms used on people currently incarcerated offers a unique analytic view to think about the dialectic relationship between data and algorithms. As we demonstrate, algorithms and data co-produce one another.

      Through a qualitative analysis of oral histories and policy documents, we show the

competing and even antithetical goals of the PACT. Finally, we investigate the outcomes of the PACT itself using a data set provided to us by the Pennsylvania Department of Corrections (PADOC). If the policy promises of a "smart" data driven approach to corrections were true, then the increased reliance on sophisticated classification tools such as PACT should quell disparate outcomes in the PADOC. To date, we find that it has not. Instead, we argue that the PACT enacts and cements the lives of incarcerated people through the enactment of data. Our case and its emphasis on what happens to people while they are incarcerated demonstrates the longer history of this type of encoding. Given the long history of these types of algorithmic and big data management practices in corrections, increased attention to carceral big data practices are an essential facet of a critical investigation of big data practices writ large.

Our quantitative analysis of the PACT draws from an intensive analysis of the metadata as well as bias in the algorithm to demonstrate the flaws in the algorithmic process. Since this algorithm is used only for incarcerated people, the PADOC collects the data, owns the algorithm, makes decisions based on the data and yet the data is still incomplete and of low quality. Through an analysis of the metadata procured from the PADOC, we find the algorithm, despite the relatively closed nature of the dataset, draws from incomplete and missing data to produce its output. The underlying data is only the beginning of the fallibility.

**Literature review**
*The relationship between data and algorithms in Criminal Justice*

In order to understand the power and role of algorithms in society, it is necessary to understand how they are generated by data and how they generate data. Examining an algorithm embedded in the carceral experience is an ideal algorithm from which to gain this vantage point. Predictive algorithms are historically rooted in the century-long scientific rationalization of and for criminal justice. Sociologist William H. Burgess as early as 1928 asserted predictability was feasible and by 1935, parole prediction instruments were being used in the US (Harcourt, 2006). In the past 30 years or so, the actuarial method has been enhanced via computational tools, and algorithms have been embraced as "a more efficient, rational, and wealth-maximizing tool to allocate scarce law enforcement resources" (Harcourt, 2006: 2). Data driven, "smart on crime" reforms almost always involve the "incorporation of dynamic risk and needs assessment into justice processes," (Schrantz et al., 2018). As the increased popularity of the smart on crime approach emerged, it is easy to lose sight of its longer history as a method of exacting justice as well as its rooted role in encouraging predictive methods across society (Wang, 2018). Even in this larger context of such methods, criminal justice in the United States remains deeply problematic. The modern criminal justice system is replete with bias against the poor and people of color in every node from arrest (Beckett et al., 2006), to sentencing (Starr, 2014), and then parole (Winerip et al., 2016).

Although the actuarial approach to criminal justice is concurrent with a justice system proven to be overwhelmingly ineffective and replete with bias, there remains a deep underlying faith among practitioners and policy makers that dynamic models can reduce costs, quell human bias, and maintain public safety (Braga, 2017; Ramachandran, 2017). Reliance on data driven techniques by policy makers continues and their implementation is presumed to make the criminal justice system more efficacious and more just, inherently (Harcourt, 2006; Završnik, 2019). Yet, structural inequity remains. In our analysis of the PACT, we reveal how the algorithm itself enacts, cements and entrenches bias through its dialectical relationship to data. The PACT offers this vantage point precisely because of this situatedness. To date, there is a great deal of attention to algorithms in criminal justice, but this overwhelmingly attends to sentencing and parole. There is hardly any study of algorithms and data that iteratively produce and control a person's experience *while* incarcerated. In any case, little attention is paid to the cycles of data that undergird the algorithms.

Critical data studies offers a lens to consider the data because it situates (big) data in time and place (Dalton and Thatcher, 2014). No data set exists in a vacuum and "the relationality of big data is far from a neutral arbiter of technological 'progress'" (boyd and Crawford, 2012; Pickren, 2016). Doursh and Gómez Cruz (2018), attuned to the social and historical nature of data, call for an increased attention to the narration of data. Data do their work in relational ways: in relation to systems of processing and in relation to people. Data also work in relation to algorithms. Algorithms are often produced by data and they create data. Carceral algorithms create data in two important ways. First, they generate additional data points via their output. Secondly, as the PACT exemplifies, they determine someone's future position and thus structure the human behavior that will be further captured as data. Building on this point in her investigation of carceral capitalism, Jackie Wang cautions, "predictions...*enact* the future" (Wang, 2018: 48 emphasis original). Barnes (2013) notes the data encourages us to look at what *is* rather than imaging what could be, highlighting the inherent tendency for algorithms to promote and even entrench the status quo explicitly through their relationship to data.

In as much as society is racially coded, so too is the data, as that is the inherent context of a racially disparate prison system. Algorithmic improvements are "sold as morally superior because they purport to rise above human bias, even though they could not exist without data produced through histories of exclusion and discrimination" (Benjamin, 2019: 10). The historical and situational context of the data that informs the PACT reveal it as rife with systemic biases due to the data upon which it relies. This creates what is functionally a feedback loop in which predictive tools enact the future data set and then "enshrine bias because they use datasets that are themselves tarnished by racial bias" (Wang, 2018: 50). Such scholarship demonstrates the propensity of data-driven criminal justice to reinforce pervasive inequities rather than alleviate them. Scholars identify new algorithmic technologies in myriad systems that, upon closer investigation, "reflect and reproduce existing inequalities," even though they "are promoted and perceived as more objective or progressive than the discriminatory systems of the previous era" (Benjamin, 2019: 5–6). Scholars have found such entrenchment by obscuring in calculations ranging from credit scores (Nopper, 2019), predictive policing (Shapiro, 2019), and insurance (O'Neil, 2016). While a great deal of work examines bias within the algorithm, there is little of this work examining carceral algorithms and the underlying data that informs them.

Working at the interface of a critical approach to data and to criminal justice calls into question the production of the data itself. Specific to criminal justice, there are myriad examples that call into question the objectivity of an algorithm because the data itself is not objective. A *New York Times'* report on the "Scourge of Racial Bias" highlights the racial biases encoded in prison disciplinary apparatuses. Of the 59,354 disciplinary hearings in New York State prisons in 2015, inmates only won about 4 percent of the time (Schwirtz et al., 2016). This study connects the production of data points about incarcerated people in New York to the pervasive racism amongst corrections staff. In one New York prison with 998 correctional officers only one is Black. The bias is indicative of a geography of residential segregation in which, "The guards who work these cellblocks rarely get to know a Black person who is not behind bars" (Schwirtz et al., 2016). The data points are produced in a field of racialized surveillance.

The study of NY State Corrections gives context to the data input in an algorithm while ProPublica finds software used across the United States to predict criminal behavior is also biased against Black people (Angwin et al., 2016). Despite these well documented concerns specific to criminal justice, predictive algorithms continue to play a central role in the US criminal justice system for pretrial and bail decisions, criminal sentencing, probation and parole, and juvenile justice (Christin et al., 2015). In their analysis of criminal justice algorithms, Christin et al. (2015), note "Instruments such as LSI-R and COMPAS are used for many purposes, including the security classification of prison inmates but also their eligibility for parole and levels of probation and parole supervision." Nevertheless, all of these algorithms produce additional data points that inform other algorithms at different nodes of the system.

Moreover, the algorithms tend to request and produce data that prioritizes incapacitation above rehabilitation (Harcourt, 2006). If rehabilitation were the primary driver, the data collected and produced would likely be quite different. Ironically, at the same time, data that could be used to evaluate the effectiveness of algorithms and implement meaningful reforms does not exist: "For example, in addition to predicting risk and providing recommendations about sentencing, algorithms could compare the actual decisions of the judges and prosecutors who used them and compile comparative statistics at the level of the jurisdiction" (Christin et al., 2015). Questioning both content and type of data can be a step towards implementing software tools that would address bias (Green, 2020).

*Data issues*

A full problematization of algorithms requires a deeper problematization of data both analytically as discussed above and technically as this section outlines. It is necessary to explore the ways in which criminal justice data management is rife not only with biased data but also with incorrect data. Logan and Ferguson (2016) highlight how fallible data in the hands of government decision-makers has negative impacts on individuals (loss of employment, unlawful arrests, etc.) and on the government systems themselves (loss of trust from the governed, inefficiencies in the system, etc.). In fact, according to a 2001 report from the Bureau of Justice Statistics, "[i]n the view of most experts, inadequacies in the accuracy and completeness of criminal history records is the single most serious deficiency affecting the Nation's criminal history record information systems" (Belair et al., 2001). In this project, we seek to explore the specificities of the concern in the context of the PADOC and how it informs the PACT.

Ferguson (2017) discusses a framework for data quality in predictive policing that also applies to our attempts to understand the PACT and PADOC data. He distinguishes between data errors that result from human error and from data that is fragmented and biased. In any large scale dataset generated by human activities and collected and catalogued by humans, there is ample opportunity for human error. Additionally, criminal justice data is being collected by a variety of organizations at a variety of levels (state, local, national). Illogical birthdates, incorrect addresses, and so on are common. The quality of criminal history records, a key component to prison decision-making, is not immune to human error, either. States have a backlog of unprocessed disposition forms, rap sheets have been shown to contain incorrect information, gang and sex offender registries have been shown to be error-prone in addition to being biased.

Ferguson (2017) also identifies methodological concerns in the way complex statistical tools are used in the decision-making process. Namely, he identifies internal validity, external validity and error-rates as three important concerns. Internal validity is the extent to which a piece of evidence supports a claim of cause and effect. Threats to internal validity such as history, change in instrumentation, selection bias are all present in the context of data-centric criminal justice. External validity is the extent to which conclusions reached in one context can be applied to other contexts. Due to the fragmented nature of criminal justice data, there is a great deal of overgeneralization that goes on. Conclusions reached in one context (the current governor of the state, the current director of the department of corrections, the current national attitude towards the carceral system) maybe should not be reached in another. Error rates are measures of the false positives and false negatives in any decision. Before asserting that an algorithm is biased, it is first necessary to establish the veracity of the underlying data.

We look specifically at the Prisoner Additive Classification Tool (PACT) used by the Pennsylvania Department of Corrections to do this. This paper highlights the role of data in criminal justice algorithms by analyzing the PACT as both data and algorithm. Algorithms permeate decision making at each intercept of the criminal justice system. In turn, the algorithm used at one intercept, produces a data point that informs an algorithm at the next intercept. The PACT is an algorithm used by the PADOC during prisoner intake and recalculated annually. The numerical score the PACT calculates equates to a custody level within the prison. It also represents a data point that informs other predictive algorithms used at other links in the system. The original purpose of the PACT is not to determine parole decisions but PACT scores determine one's life in prison. Subsequently the PACT is not only a variable used in parole algorithms "upstream," it enacts the future data points of the incarcerated person. We thus find it necessary to consider algorithms and data in dialectical interplay. In what follows we contextualize the PACT in a wider context of PADOC priorities, we then look more closely at the issues with the data used by the PADOC to calculate it, and we finally look at biases embedded in PACT calculations as problematic future data points.

**Case study: History of the PACT**

*The PACT in Historical Context*

The PACT is an algorithmic reform implemented by the PADOC in 1991. The tool's implementation exemplifies Pennsylvania's position as a vanguard of national prison reform. Home to the first modern penitentiary, Eastern State Penitentiary in Philadelphia, Pennsylvania has long been a national leader in correctional reforms and innovations. When the state introduced the Pennsylvania Additive Classification Tool in 1991, it was regarded as a leader in nationwide efforts to create an "objective" prisoner classification system (NIC, 2015). The Justice Reinvestment Reform Initiatives of the past decade derive from the same techno-moralism that lead to the implementation of the PACT. This place-based context adds to the ways the PACT is a unique carceral algorithm to study. It is exclusively for incarcerated people: its only inputs are from the DOC meaning that it is theoretically an ideal system for a complete data set. The PACT as indicated in Fig. 1 is a confidential algorithm. The resulting custody level and other input data we examine was requested from the DOC, and it is not publically available. Since it is an algorithm for prisoners, it is discursively imputed with expectations of being free from any type of bias while simultaneously being explicitly driven by population control concerns. This concern, as we discuss, is in and of itself a bias – that is to say attention to data production invites a necessarily broader conception of bias.

The PACT was implemented as part of a sweeping set of reforms to deal with the exponential growth of the PA prison population in the 1980s. This growth had reached a tipping point on October 25, 1989 when an uprising began in the Camp Hill State Correctional

Institution. This disruption, now known as the Camp Hill Riots[1], lasted for three days. When the literal dust settled, "more than 100 people were injured including 69 corrections officers and 41 inmates" (Kiner, 2018). It resulted in over $50 million dollars of damage to the facility. Over the next year, additional disruptions happened in several other prisons across the state (Stanford, 2005). The Camp Hill Riots, in as much as they symbolized the need to reconceive prison management in the wake of a burgeoning population, played a notable role in reshaping PADOC policy. This impact was compounded because Camp Hill also houses the central offices of the Department of Corrections and the disruption was, therefore, witnessed by decision makers that do not have day to day experiences within the walls of the prison.

The commemoration and reflection efforts 30 years later underscore the significance of the Camp Hill Riots for shaping the PADOC and give a unique window into the historical context that underlies the implementation of the PACT. In 2019, to commemorate the 30th anniversary of the Camp Hill riot, the Pennsylvania Department of Corrections Communications Director Susan McNaughton conducted 43 oral history interviews with people who were PADOC employees during the riots[2]. There is much that these oral histories reveal about the incident and the PADOC. This compendium, to its credit, reflects the data-informed, transparent, and reflexive tendencies of the PADOC in its willingness to recall and review the incident even 30 years later. Such tendencies have made the PADOC a leader in corrections across the country. The compendium accounts also reflect a sensibility that the PADOC is now "fixed" as it is currently well equipped to manage a much expanded population. It is considered beyond the purview of the PADOC to question the population size constructed by legislatures in the first place.

In this compendium, which explicitly notes the development of new classification tools to manage and classify incarcerated people, we see the gaps in control and other system deficiencies that motivated the implementation of the PACT. Our analysis specifically finds the emergence of security and technological innovation as values that emerge from the riots and guide subsequent reforms like the PACT tool. The lived experience of the riot for those at the PADOC central office had indelible impacts on their own understanding of the prison and the expanding incarcerated population. Many of the PADOC personnel interviewed who witnessed or experienced the riot, recollect traumatic scenes that still impact them thirty years later:

> "There was an officer, in particular, that was on his knees… right inside the gate below the tower… that had a pillowcase over his head or his head was covered. There were about three inmates with him that were swinging a shovel or a pick-axe or something. I just remember being absolutely sick to my stomach that we were going to watch somebody killed directly in front of us. Frank, in particular, was very vocal… very upset that the officer in the tower did not fire a shot. It seemed like it went on for 20 minutes, until finally he did fire a shot. I think I recall Frank running out and telling him to, "Fire! Fire!" I just was only a 26-year-

---

1 These incidents are widely known and referred to as riots by the Department of Corrections and the news media coverage of the incident. We refer to them as the Camp Hill riots throughout for clarity and deference to colloquial terminology while acknowledging the words severe limitations (Wang, 2018).

2 The interview transcripts range in length from 1441 words to 9229 words and the total word count of the entire corpus is 196508 words. Using the Voyant-Tools software, we conducted a critical discourse analysis of the interview transcripts and the additional documentation published by the DOC commemorating the Camp Hill Riots.

old kid at the time, but that moment is seared in my brain. When I think back to the riot… just that… the inmate swinging that shovel and then laughing. I don't think I'll ever get rid of that picture in my mind." - Lee Ann Labecki (2019), October 3, 2019, Pennsylvania Department of Corrections Employee Oral History Collection Project

"Bickell: At that point, there was a lot of threatening going on. A lot of punches. They handcuffed us. My partner had thick glasses – I remember they took his glasses and they crunched them, they just stepped on them, so he couldn't really see anything. But they wound up separating us. I was handcuffed – I was kind of like hog tied. I got handcuffed pretty much. Originally, I was handcuffed with my hands and my legs together almost – you know – not a good position to be in. So, they did that and there was a lot of threats being made.

McNaughton: Did they blindfold you?

Bickell: Yes. They covered me up then, and then a couple inmates that were on the block with us... they came up and asked if we were ok and I told them the handcuffs were tight. So, they got me from that real vulnerable position and then just put me in the regular handcuffs. Then they loosened them up. And then they just started walking us around. Again, I eventually met up with other hostages – I can't remember who they were – I remember, but I can't, if that makes sense." Tabb Bickell (2019), Pennsylvania Department of Corrections Employee Oral History Collection Project

These quotes are just a few that exemplify the lingering intensity of the experience for corrections staff. The words "scared" or "fear" appear in more than one third of the transcripts. While these emotions are not always explicitly mentioned, the interviewees describe harrowing scenes of adrenaline-fueled crisis conditions as the riot wore on for a few days.

The accounts are nuanced in as much as they recognize the different inclinations of incarcerated people and in the tacit acknowledgement of compounding factors. In reflecting on what caused the Camp Hill Riots, interviewees recognize the fundamental problem of overcrowding in prisons across the state. Leanne Labecki, the current director of Research, Planning and Statistics for the PADOC reflects on her experience of the Camp Hill riots when she was an intern in the main office:

"I remember in the months prior to the riot, in central office, we had been preparing a number of planning documents on the level of overcrowding in the system generally. And certain institutions in particular, one of them being Camp Hill, because the people that were long-termers kept saying, "It's not a matter of if. It's a matter of when."...We had some institutions, like Camp Hill and others, that were at 145% of design capacity and rated capacity we were up to 180... and they were putting beds everywhere... in dorm rooms... in gymnasiums. Ted and I were developing plans with other departmental staff on how to address crowding. We're sending things over to the Governor's Office to make them aware." - Lee Ann Labecki (2019), October 3, 2019, Pennsylvania Department of Corrections Employee Oral History Collection Project

Labecki, like many others who were working in Central Office on October 25, 1989, were shaped by the riot and their memories of it influenced the trajectory of their careers and, in turn, PADOC policy. She makes clear overcrowding and the failure of PADOC to manage the population created a tinderbox effect. The subsequent riot, a substantial loss of control, became a pivot point that so much prison policy and prison decision making is threaded through. The path dependencies of solutions to overcrowding and subsequent prisoner unrest is to build more prisons and, notably, prisons that are more sophisticated. This sophistication emerged on two fronts in the early 1990s: in the physical construction and design of prisons, and actuarial approaches to inmate intake.

The interviewees often comment on different construction issues that have since changed as they describe the way control was ultimately lost by staff. Tabb Bickell's oral history notes infrastructural improvements since the riots as he reflects on his experience at the time:

> "The switchbox was a lower area where you actually ran the – it was like a control center almost – they were levers at that point, there were not push buttons like you have now."
>
> "At that point I remember that the inmates -- and this is some of the good things that has happened since then – one key got you everywhere. One door key got you everywhere. So that once those inmates got the keys, they were coming through our side door – what we called the dayroom – they were coming from E Block, and I realized that we were in trouble."
>
> "Radios back in them days were what they were. They didn't work – not like we have today."
>
> -Tabb Bickell (2019), Pennsylvania Department of Corrections Employee Oral History Collection Project

This interview reminds us that such a loss of control today is less likely due to the massive construction and technological improvements of the PADOC built infrastructure since that time. In the hindsight narration, it is a culminating event for a system that was wholly ill-equipped to manage the new population. In 1977, the PADOC population on December 31 was 7,561 and by 1989 it had grown to 20,490. In this same time, the state saw a general population increase of only one percent (Planning, Research and Statistics Office, 1989). At the time of the Camp Hill riot, the maximum capacity for the prison was just over 1800 but there were 2500 men crowded into the facility. Like most prisons across the US, the war on drugs was the primary driver of this increase. While some interviews acknowledge the incarcerated men who participated had articulated demands made in response to overpopulation conditions, but correctional personnel primarily focus on loss of control, interviews commonly circulate around the following themes: lack of resources, technological inferiority in prison design, and little ability to manage prisoners as a population. Our analysis of the oral history interviews reveals that the word "control" is one of the most used words in the corpus, ranking 42$^{nd}$ and mentioned 157 times in the interviews outside of the phrase "control center." From the perspective of the PADOC, the crisis of the Camp Hill Riots is hooked on the loss of control overpopulation generated not the conditions it created.

After the Camp Hill Riots, the PADOC implemented an algorithmic method to sort inmates. Prisoner intake is standardized by the PADOC shortly after and in clear reaction to the Camp Hill Riots. In this period, prisoners become seen as a population to be managed. The riot marks a final shift away from, in a Foucauldian sense, the individual management of prisoners to the organized management of a prison population. Figure 2 from the Department of corrections website notes the major changes that were implemented in response to the Camp Hill Riots. The first one: "Elimination of the treatment vs. security struggle by changing to a facility management and centralized services structure," is referring in part to the development of the PACT for prisoner classification.

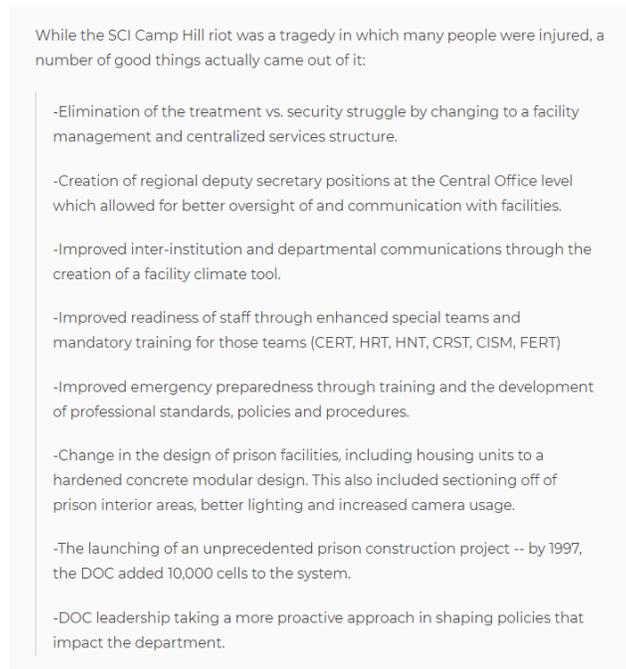

*Figure 2: Changes Since the Riot (2019)*

In total, reviewing documentation by the PADOC leading up to the implementation of the PACT, reveals the underlying orientation of the PACT: that using technological sophistication via "objective" algorithmic methods to quickly and efficiently manage the growing population, it is primarily meant to serve an incapacitation rather than a rehabilitative function. Discursively, security and control are the fundamental concerns that drive reforms in the early 1990s and the PACT is part of this effort. In a report offering the PACT as a model for other jurisdictions, the NIC celebrates the techno-sophistication of the tool noting, "It was developed by an interdisciplinary team as a risk management tool for placing prisoners in the least restrictive custody while providing for the safety of the public, community, staff, other prisoners, and institution guests and visitors and for the orderly operation of the institution" (Hardyman et al., 2004: 37). This makes clear the primary purpose of the PACT is to systematize population management to maintain control and secure the prison. In reading for absence, it is clear that goals such as rehabilitation, eliminating racial bias, and addressing community scale structural issues are not part of the equation. While the possibility of architectures, programs or variables that are oriented toward restoration and rehabilitation likely would have also quelled unrest, this is not the path pursued and that lack is embedded in the algorithm.

*The PACT today*

Still in use today, the PACT is meant to standardize the custody assignment process and to systematically sort prisoners based on a prediction of their institutional behavior. The PACT thus follows a long actuarial tradition in evaluating and structuring the lives of incarcerated people (Harcourt, 2006). The PACT number equates to a custody level where CL-1 is the least restrictive and CL-5 is the most restrictive (see Table 1).

| Custody Level | Corresponding Housing Unit Assignment | Examples of Associated Privileges |
|---|---|---|
| 1 (least restrictive) | Community Corrections Center | Can leave the facility daily for work, family visits, and programs. These are structured as halfway houses and also house parolees. |
| 2 | Minimum Custody Housing | Can participate in work release programs |
| 3 | Medium Custody Housing | Can hold more jobs within the facility |
| 4 | Close Custody Housing | Can participate in vocational training with special permission |
| 5 (most restrictive) | Maximum Custody | |

*Table 1: Custody levels (compiled using Hardymann et al. 2004 and PADOC 2011)*

These custody levels have dramatic consequences for prisoners as it will determine their "housing, programming, and freedom of movement" within the prison and even which facility a person is located in (Hardyman et al., 2004: 37). The report further discloses the way in which the PACT supposedly removes prison staff from the decision and the PACT is thus represented as a naturalized reflection of the prisoners:

> PACT was designed to be objective and behavior driven and to ensure that a prisoner's custody level is based on his/her compliance with institutional rules and regulations and participation in work, education, treatment, and vocational programming. Thus, in theory, compliance by the prisoner will facilitate his/her movement to less restrictive custody levels. PADOC discourages negative behavior by providing consequences for infractions, escapes, and nonparticipation in programs. (Hardyman et al., 2004: 37–38)

This language narrows the responsibility for one's custody level squarely onto the incarcerated person. This reflects a neoliberal algorithmic logic that obscures systemic bias (Story, 2019; Wang, 2018). Further, the stated goals of the PACT are fraught and somewhat contradictory as the documentation waffles between emphasizing the incapacitation function as well as the rehabilitative one. The NIC report by Hardyman et al. (2004) makes clear that the classification system, which is conceptualized as a technological integration to manage a burgeoning prison population, is also integral in rehabilitation efforts. The classification process as a whole also includes a needs assessment that delimits additional services and programming someone may require. The high value on technological sophistication that emerged during the late 1980s and culminates in the PACT threads its way into more recent policy reforms in the state, as does the PACT calculation itself.

      Given Pennsylvania's long tradition of actuarial policy and algorithmic logics that supposedly remove human bias, the system remains rife with bias (Jasanoff, 2017; Kramer and Ulmer, 2008; Schwirtz et al., 2016). It is difficult to fully analyze the extent and force of embedded bias in the data that feeds the PACT because the algorithm is not publically available (see Figure 1). The public facing explanation of the PACT redacts the algorithm (PADOC, 2011). A major justification for this redaction is a concern that prisoners could game the algorithm (*Raymond J. Smolsky V Department of Corrections*, 2011). This, of course, inadvertently reveals that the PACT is not as objective and free of human intervention as we are invited to assume.

<span style="color:red">**CONFIDENTIAL**

**11.2.1, Reception and Classification**

**Section 3 – Pennsylvania Additive Classification Tool (PACT)**

**This section is confidential and not for public dissemination.**</span>

*Figure 1: Redacted section of the Reception and Classification Manual that contains the PACT algorithm*

This vulnerability to manipulation along with consideration of the social and historical factors that led to the development and deployment of PACT indicate that the PACT is fallible and objective. A closer examination of how PACT scores are overridden in practice by corrections staff further problematizes the notion of an objective and accurate carceral algorithm. The NIC report cited above notes the use of overrides that are used to reclassify prisoners after calculating the PACT score and generating a custody level:

> Administrative overrides—based on the prisoner's legal status, current offense, and sentence—can change classification recommendations. Discretionary overrides by the case manager are permitted based on the prisoner's security threat group affiliation; escape history; nature of current offense; and behavior, mental health, medical, dental, and program needs. Information about cases for which discretionary overrides are recommended is forwarded electronically to the appropriate staff for approval. Multiple levels of review by classification supervisory staff and the central office are required for all overrides. (Hardyman et al., 2004)

This is a tacit acknowledgement that the PACT does not always classify "correctly", necessitating a human intervention. Our historical attention to the PACT shows the underlying priorities that are embedded in the technology itself. Addressing racial bias is not mentioned anywhere in the documentation, if anything it is further obscured by the implicit assertion that human agency (i.e. human bias) is reduced.

In spite of these signals of complexity, what emerges is an assumption by state officials that through increased efficiency, analysis with more data points, and models making decisions,the scourge of biases and injustices are inherently eliminated at a cost savings to taxpayers. The impetus to be "data driven" firstly implicates the use of data in the creating and justifying of reform policies. Data driven decisions help legislatures calculate savings and redirect resources in presumably more efficacious ways. The faith in an enhanced prison through data driven decision making has proliferated the governing logic of the institution. Take, for example, the case of a concerned citizen in a town hall facilitated by the PA Legislative Black Caucus. The citizen asserted to PA Department of Corrections Secretary, John Wetzel, that incarcerated people were not receiving sufficient portions of food. Wetzel offered the following response:

> The Department of Corrections' Master Menu is analyzed using computer software. The DOC menu, as written, when served the standard portion sizes as indicated on the menu, meets the current Recommended Dietary Allowances and Dietary Reference Intakes, Males and Females, 18 to 50 years, as identified by the Food and Nutrition Board of the National Research Council. Every facility uses measured portion utensils, pre-portioned items, or scales to ensure the standard portion is served. Many people are not used to managing or limiting their food portions to those recommended for a healthy lifestyle. It is our hope that inmates can learn what healthy portions look like so they can in turn carry that forward once released (Pennsylvania Legislative Black Caucus, 2018).

The dismissal of the question at hand captures the ways data driven decision making presumes morality, fairness and goodness. It obscures the larger questions of justice and human rights through a revealing of the data that drives the food distribution system and reinforces the individualized deficiencies of incarcerated people who are hungry. Given the rightness of the *data*, anyone who is hungry is surely wrong. This reliance on data to drive decisions and the subsequent belief that the decisions are objectively right has the potential to intensify injustices. Implementing more data in each phase of corrections assumes a better prison is being constructed, without ever unpacking the existence of the prison itself let alone the data.

**Data driven Methods and Results**

The historical context of the PACT foregrounds the actuality of the PACT calculation by the PADOC and its impact on incarcerated people. In what follows, we analyze the data that the PADOC collects and uses in the algorithm. The nature of the data set and a summary of the demographics of inmates over a 25 year period give a brief window into the structural inequality in the PA prison system. Following, we discuss the decision making process employed to determine an inmate's classification or reclassification scores and point out some issues both with the algorithm itself and the data used to run the algorithm. Finally, we highlight biases in the outcomes of the PACT that persist to this day, despite the insistence that "smart on crime" tools are more objective and fair.[3]

*Nature of the dataset*

In July 2018 we requested a data pull from the PADOC using the Pennsylvania Department of Corrections Research Approach Request Form (RARF)[4]. The intent of the request was to study factors that influence parole decisions. The PACT is one such factor and we found it captures a significant node in a person's flow through the correction system as visualized in figure 4 below. We requested data on incarcerated people, including those who have been paroled, who were in the system in 1997, 2002, 2007, 2012, and 2017. We requested variables that were related to parole decisions. Nine months after our initial request we received access to data on more than 280,000 distinct incarcerated people; of those only 146,793 were incarcerated in the years we requested.

The data was composed of 20 files and an accompanying data dictionary and code book. The data dictionary clearly indicated that the DOC maintains even more files than these 20 and consequently collects more data than the several gigabytes we received. The expansive data being collected, some of which we were given access to, records details spanning almost the entirety of an inmate's life in and outside the prison ranging from their date of birth to any notes recorded by their therapist during a treatment session. A good example of the reach of the data is the fact that, according to the data dictionary, the PADOC is collecting data about the criminal history of anyone who visits an inmate.

Figure 3, summarizes the racial demographics, sex demographics, and age distribution of the people for whom we have data in each of the 6 years that we requested in the original data pull. Note that recently the population has trended younger, the ratio of the number of male incarcerated people to the number of female incarcerated people has remained stable and the ratio of the number of Black incarcerated people to the number of Whites has changed from roughly 3:2 in 1997 to approximately 1:1 in 2017. This disparity remains in spite of Black people making up only about 11% of Pennsylvania's general population.

---

3 See the (cite Gitlab site) for other figures and tables related to the bias in the numbers produced by the PACT and for the code we used to produce these graphs and tables.

4 https://www.cor.pa.gov/About%20Us/Statistics/Documents/Research%20Approval%20Request%20Form.pdf

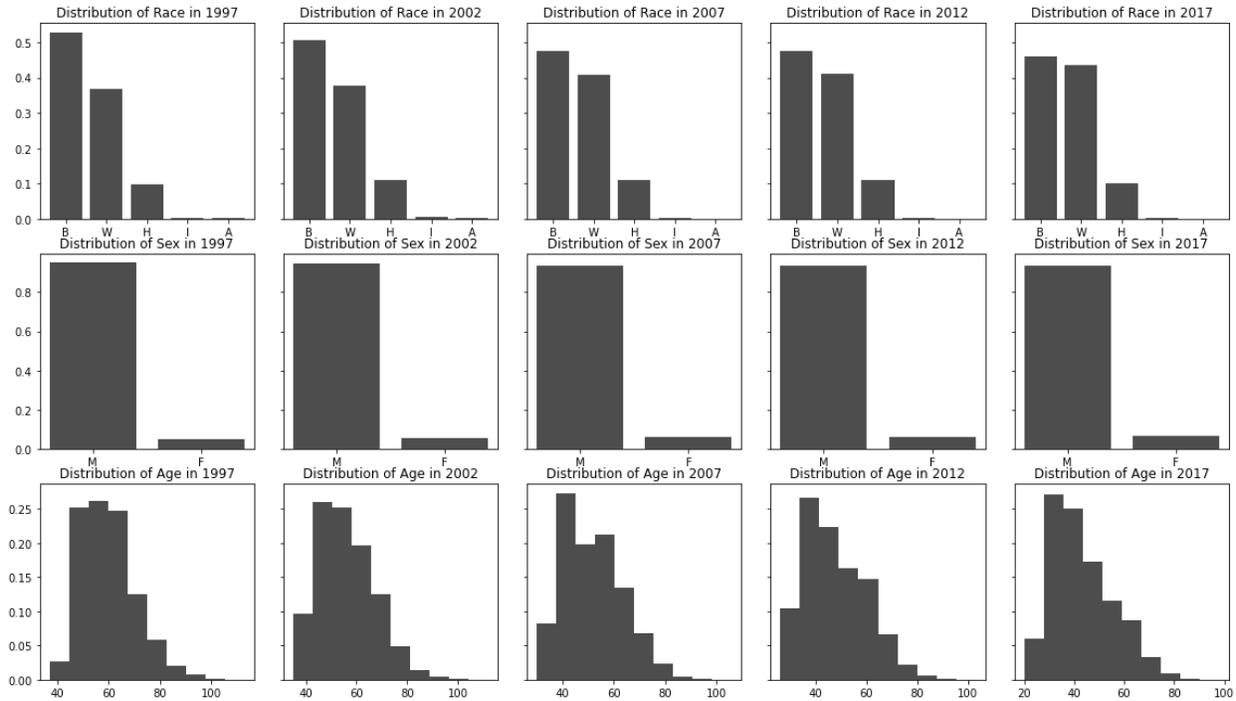
*Figure 3: Basic demographic data for incarcerated people in the system in the years 1997, 2002, 2007, 2012, 2017. Race and ethnicity categories are by first letter and correspond with figures 10 and 11.*

*Bias and human subjectivity in both the data and the algorithm in the data pipeline*

A hallmark of arguments in favor of "smart" tools in decision making is the claim that the tools are immune to subjectivity and human bias. As we note above, bias in criminal justice algorithms writ large is well established. However, this case study offers unique documentation of this bias *within* the prison. Given the power of this carceral algorithm to enact the life of an incarcerated person, the documented bias we find is especially troubling. The PACTs situated position as both data and algorithm further offers a unique window into the way bias is cemented in both data and algorithms at various nodes of the system.

Based upon a careful reading of the available policy documents and legal cases (Hardyman et al., 2004; PADOC, 2011; *Raymond J. Smolsky V Department of Corrections*, 2011; Stanford, 2005), we developed a flowchart for the process that assigns an inmate their initial classification level; see Figure 4. The process has four main steps: the PACT tool is applied and an initial score is derived (we do not know how this score is calculated or what ranges it can take on). That score, which reduces the infinite variables of a human life, highlighting their interactions with police and the courts, is then summarized as an inmate's initial classification score, an integer from 1 to 5. That initial classification score can then be overridden for administrative reasons and then a custody level score is then determined. This custody level score determines a great deal about an inmate's experience in the system. The flowchart in Figure 4 is color coded as follows: Grey boxes represent numbers whose calculation is either explicitly confidential or generally mysterious and obscure. Blue boxes represent data about an individual that are known to be biased (e.g., since Black men are known to be over-policed and get harsher sentences, we know there is a bias in those two inputs to the PACT tool). The green box corresponding to stability factors encompasses features that seem notably subjective and value laden as they include things like an inmate's marital status, employment status before being incarcerated, etc. The resulting diagram suggests that the system is neither objective nor free from bias.

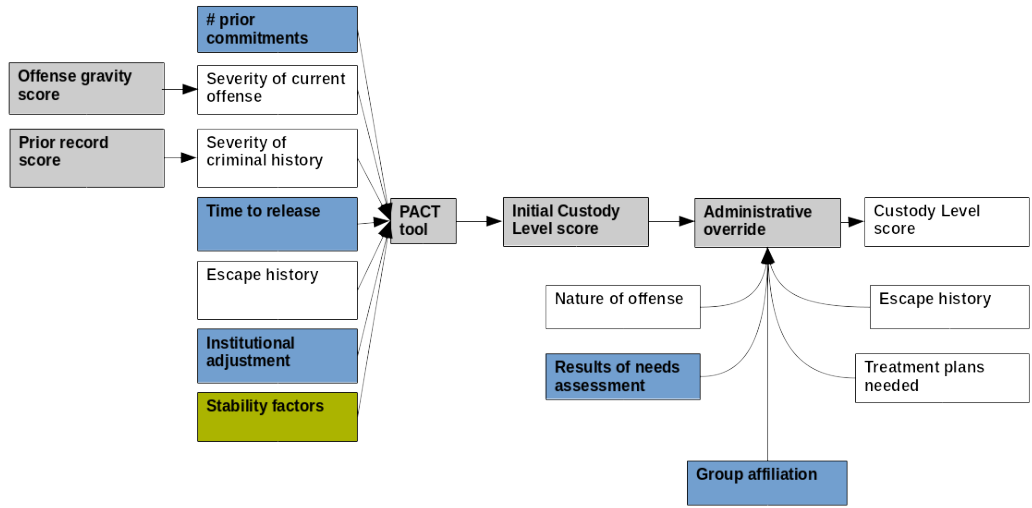

*Figure 4: A flowchart of the process to assign an incarcerated person their initial custody level. A similar process holds for the annual reclassification*

The process outlined in Figure 4 results in one of five categories of custody level[5] being assigned to an inmate. Figure 5 shows the distributions of custody levels assigned to inmates during the initial classification, after the initial classifications' administrative overrides, during an annual reclassification and their administrative overrides. The initial classification distribution is somewhat symmetric. After the administrative overrides are completed, inmates tend to have a lower custody level and there is a wider range of custody levels. Reclassification scores tend to lower an inmate's custody level but also simultaneously increase the range of custody levels.

---

5 There were several files with variables corresponding to an inmate's custody level in the data provided to us by the PADOC. The advice given to us by the PADOC was that "for many inmates you would need to combine all 4 custody level fields and take the one that is populated for the CL." And so that is what we did.

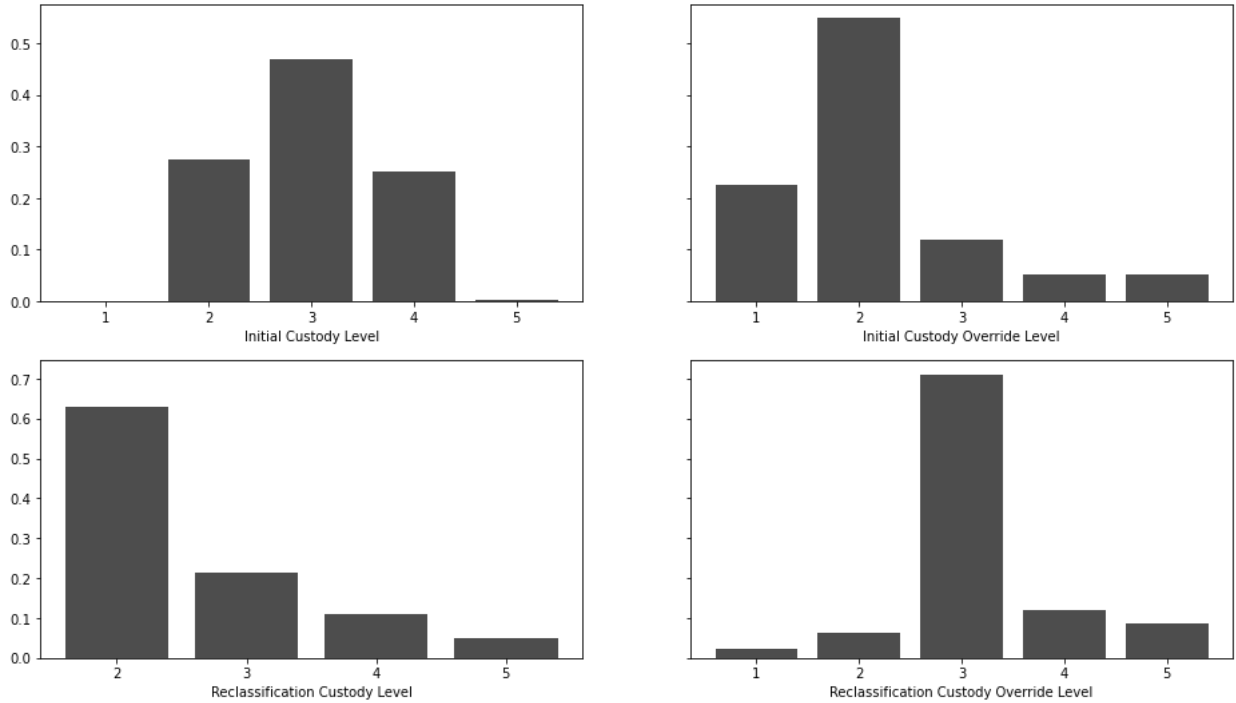
*Figure 5 Distribution of the output of the PACT and the override process for incarcerated people in the system in the years 1997, 2002, 2007, 2012, 2017..*

*Variables being collected*

      The data dictionary that accompanied the dataset we were given contained information about all the variables the PADOC collects including the date each variable came into existence. Figure 6 illustrates the growth in the number of variables starting in 1989, the year that the "smart" approach to crime became cemented in the processes carried out by the PADOC. Currently the PADOC collects data for more than 1200 variables for each inmate and the number of variables being collected for each incarcerated person increases over time. Such information about the data being collected about incarcerated people has not previously been published or analyzed.

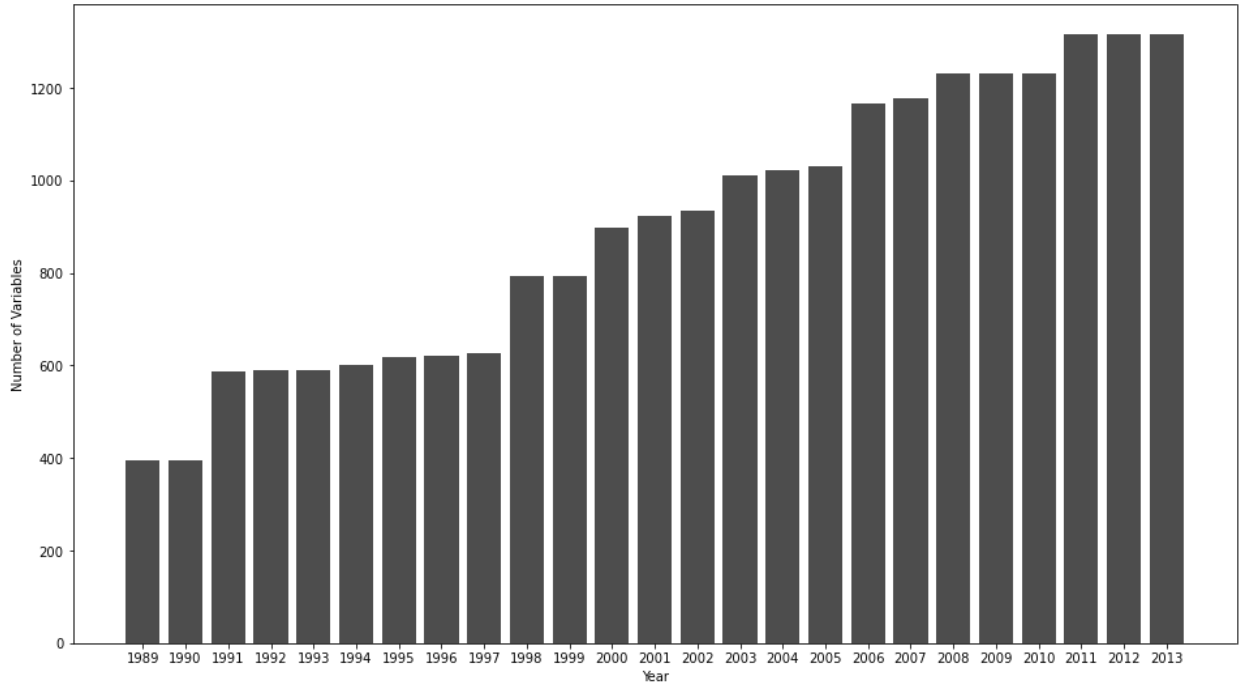
*Figure 6 Cumulative number of variables collected by the PADOC data set by year*

       Though a lot of data is being collected about each inmate, the data itself is problematic in the ways described above. Despite the PADOC's appraisal of the quality of their data (see Figure 7), there is a lot of missing data and a number of instances of illogical values[6].

---

6 For example, a variable for maximum court sentence in months should range from 0 to 12 but some values greater than 100. For example, marital status, one of the stability factors in Figure 3 should take on the values DIV, MAR, SEP, SIN, UNK, WID but we see that there 20 different values taken on including clear typos like WED and WIS that are meant to be WID and some values that are not defined anywhere. There are not many instances of each of these values, but they exist nonetheless.

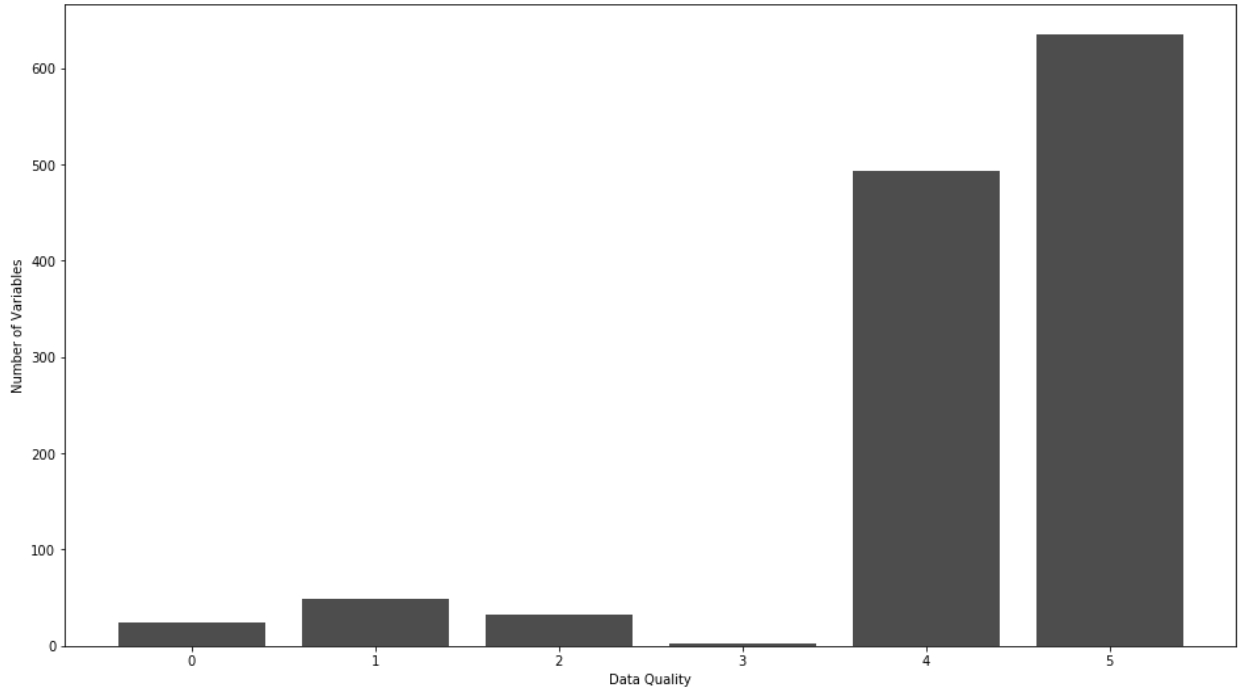
*Figure 7 Quality of variables added to DOC: 5 is the highest quality and 1 is the lowest.*

In Figure 8, we see an increase in the sheer amount of data PADOC is collecting system wide about the people it incarcerates. This is measured in the number of cells in the database that could be filled in. It is essentially the product of the number of inmates and the number of variables. Also illustrated in Figure 8 is that there is roughly an equal number of empty cells as there are cells with values. In a real sense, about half of the possible data is missing. If we focus our attention on those variables that are used in the making of important decisions (e.g., custody level score, parole), there is a great deal of missing data there, too. In Figure 9 we see that the custody level score, the output of the PACT tool, is missing for a large number of inmates. The tool is supposed to be "smart" but it does not appear to record what it knows. This offers an analysis that exclusively focuses on incarcerated people and their experiences within the prison – not just intake and release.

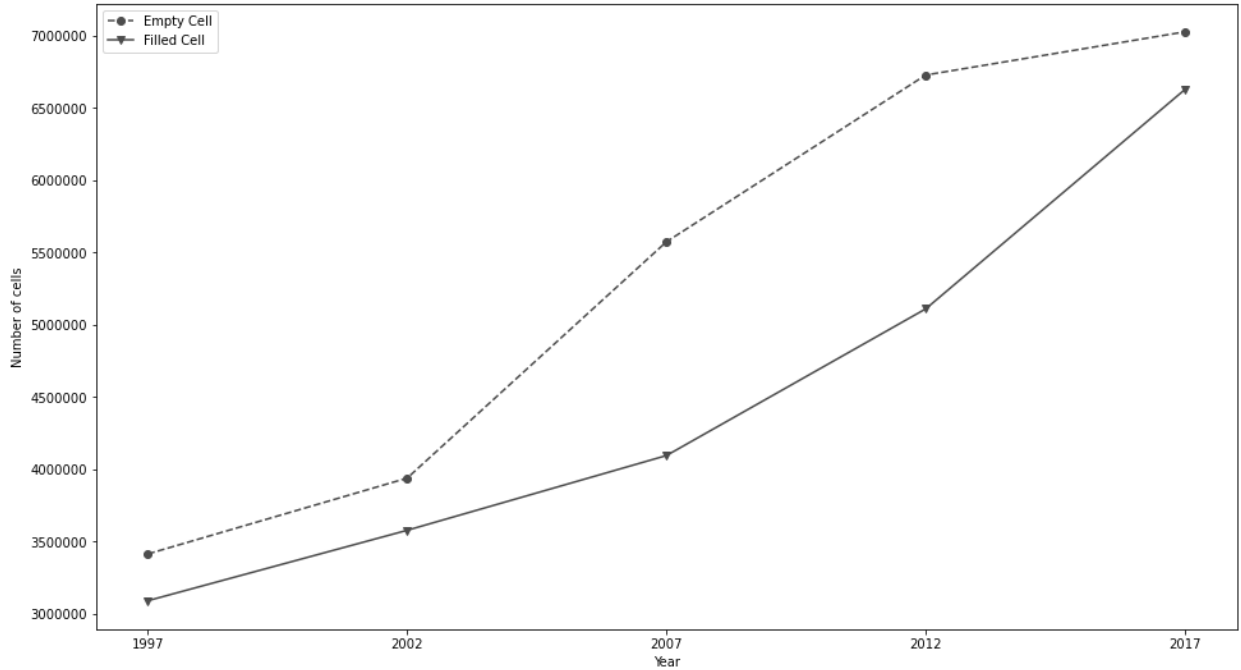

*Figure 8 Total number of empty and non-empty cells over time.*

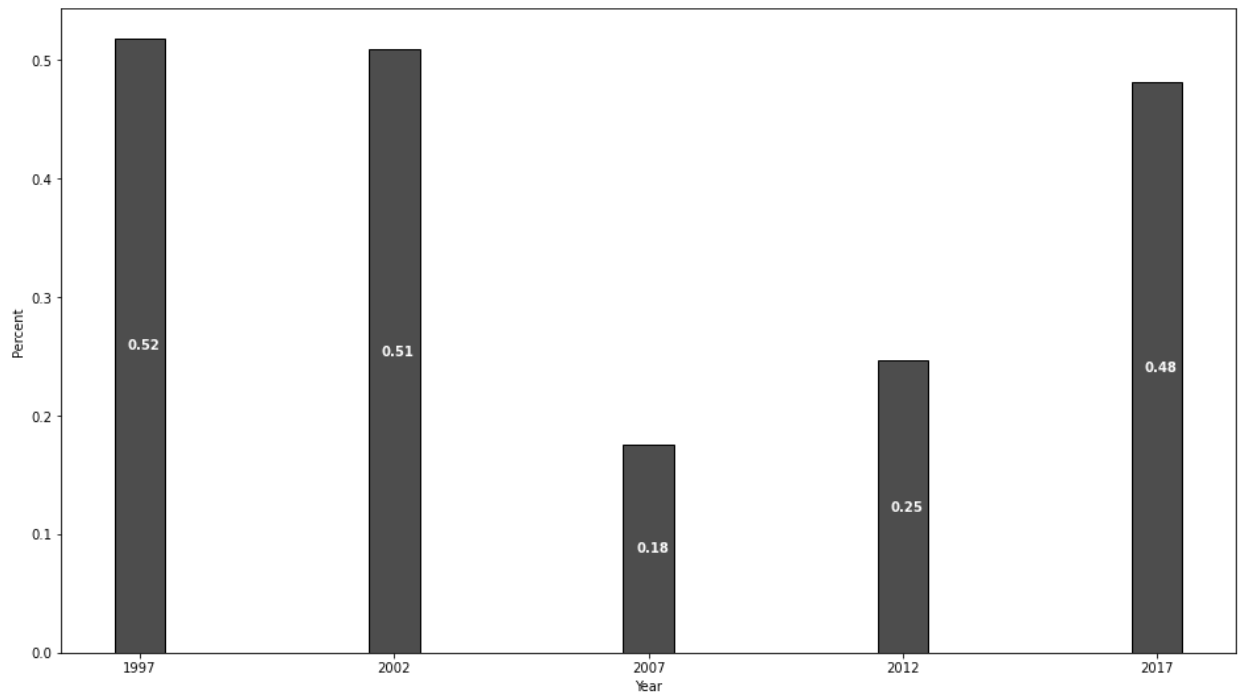

*Figure 9 Missingness of the initial custody level. The shaded areas are the proportion of cells that contain data.*

*Unsuccessful in reducing bias*

    As discussed above, "smart" tools are marketed as being able to remove or reduce biases in decision making processes. These tools have now had 30 years to achieve this goal.

Collectively, Figures 10-11 and Tables 2-3 show, this is not the case.[7] In Figure 10 and Table 2, about 84% of White inmates have their custody level reduced by 1 or more points for administrative reasons whereas the same can be said for only 79% of the Black inmates. This discrepancy suggests different treatment for the two different groups. Moreover the fact that 13.3% of Black inmates have their custody level reduced by two points or more while the same can be said for only 8.5% of White inmates suggests that perhaps the PACT tool assigns higher scores to Black inmates than it does to White inmates. These biases will be returned to in future papers.

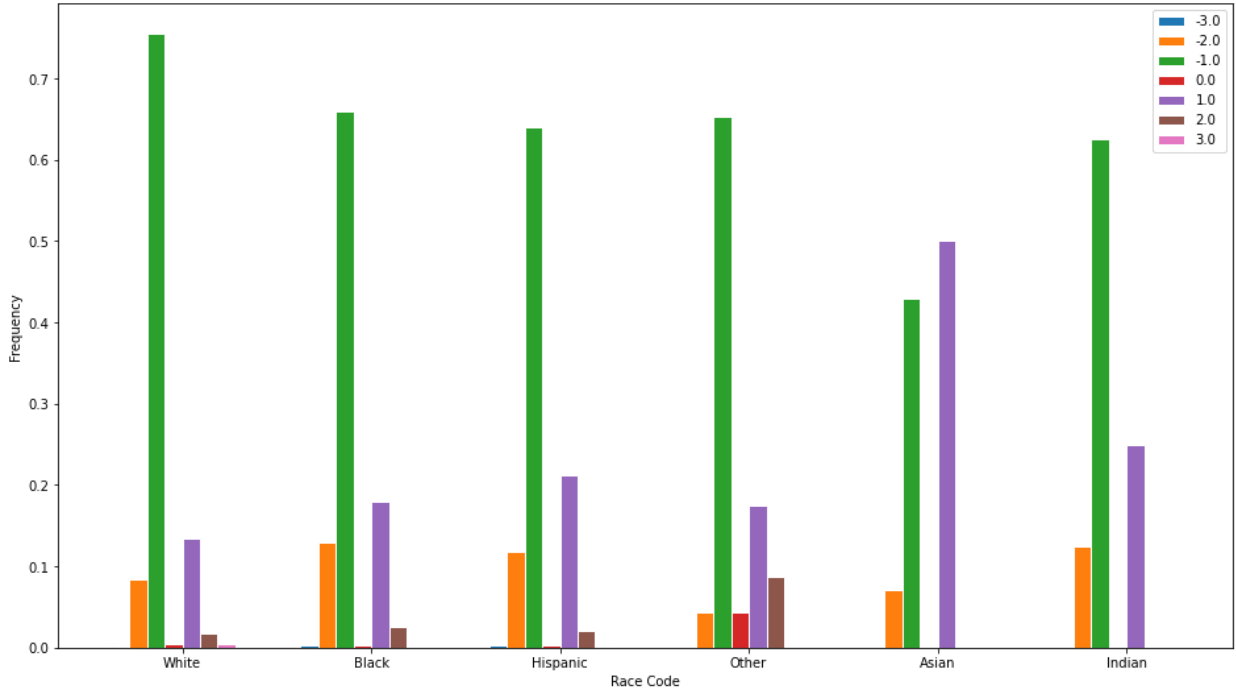

*Figure 10 Distribution by race of the difference of initial custody level and override custody level. The data is from inmates incarcerated in the years 1997, 2002, 2007, 2012 and 2017.*

---

[7] In the tables and figures in this section, we report data for fewer than 10,000 inmates even though our data set contains information for over 280,000 inmates. Since this is all of the data available we consider it a population and there is no way to know if there is any bias in the inmates for whom we have the required data, we are not performing inference.

*Table 2 Difference between Initial Custody Level Override and Initial Custody Level by Race. The data is from people incarcerated in the years 1997, 2002, 2007, 2012 and 2017.*

| Difference in custody | Race Code | | | | | | Total |
|---|---|---|---|---|---|---|---|
| | White | Black | Hispanic | Other | Asian | Nat Am | |
| -3 | 5 (0.1%) | 9 (0.4%) | 2 (0.3%) | 0 (0.0%) | 0 (0.0%) | 0 (0.0%) | 16 (0.2%) |
| -2 | 283 (8.4%) | 330 (12.9%) | 73 (11.8%) | 1 (4.3%) | 1 (7.1%) | 1 (12.5%) | 689 (10.5%) |
| -1 | 2535 (75.5%) | 1687 (66.0%) | 395 (64.0%) | 15 (65.2%) | 6 (42.9%) | 5 (62.5%) | 4643 (70.6%) |
| 0 | 13 (0.4%) | 6 (0.2%) | 2 (0.3%) | 1 (4.3%) | 0 (0.0%) | 0 (0.0%) | 22 (0.3%) |
| 1 | 451 (13.4%) | 460 (18%) | 131 (21.2%) | 4 (17.4%) | 7 (50.0%) | 2 (25.0%) | 1055 (16.0%) |
| 2 | 57 (1.7%) | 64 (2.5%) | 13 (2.1%) | 2 (8.7%) | 0 (0.0%) | 0 (0.0%) | 136 (2.1%) |
| 3 | 13 (0.4%) | 1 (0.0%) | 1 (0.16%) | 0 (0.0%) | 0 (0.0%) | 0 (0.0%) | 15 (0.2%) |
| Total | 3357 | 2557 | 617 | 23 | 14 | 8 | 6576 |

In Figure 11 and Table 3 Differences between Reclassification Custody Level Override and Reclassification Custody Level by Race, we observe something similar in the case of reclassification. 25.6% of Black inmates receive an administrative override to lower their custody level score whereas the same is true for 30.8% of White inmates. Also, 21.1% of White inmates had their reclassification overridden to a score two points lower whereas the same is true for 12.0% of Black inmates. This suggests that perhaps White inmates are being reclassified at a higher level than Black inmates relative to where they should be according to administrative requirements.

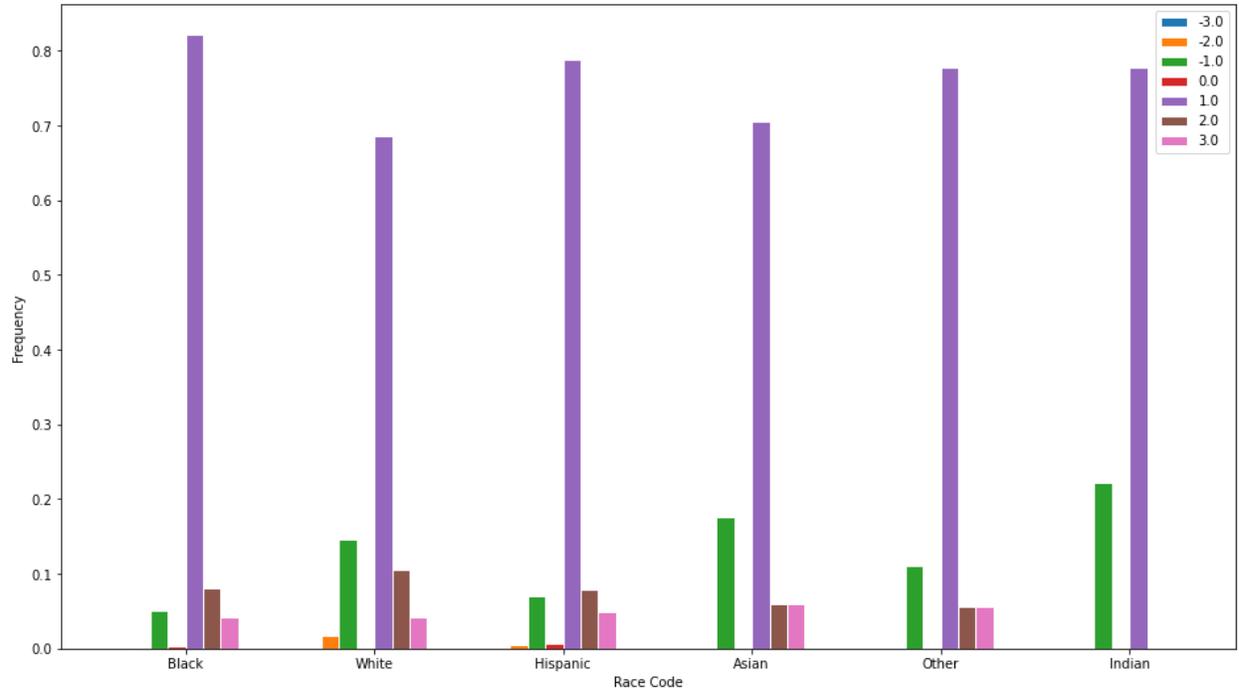

*Figure 11 Distribution by race of the difference of reclassification custody level and override reclassification custody level. The data is from people incarcerated in the years 1997, 2002, 2007, 2012 and 2017.*

*Table 3 Differences between Reclassification Custody Level Override and Reclassification Custody Level by Race. The data is from people incarcerated in the years 1997, 2002, 2007, 2012 and 2017.*

| Difference in custody | Race Code | | | | | | Total |
|---|---|---|---|---|---|---|---|
| | White | Black | Hispanic | Other | Asian | Nat Am | |
| -3 | 1 (0.1%) | 1 (0.0%) | 0 (0.0%) | 0 (0.0%) | 0 (0.0%) | 0 (0.0%) | 2 (0.0%) |
| -2 | 25 (1.7%) | 6 (0.2%) | 2 (0.5%) | 0 (0.0%) | 0 (0.0%) | 0 (0.0%) | 33 (0.7%) |
| -1 | 217 (14.7%) | 147 (5.0%) | 30 (7.1%) | 2 (11.1%) | 6 (17.6%) | 2 (22.2%) | 404 (8.2%) |
| 0 | 3 (0.2%) | 10 (0.3%) | 3 (0.7%) | 0 (0.0%) | 0 (0.0%) | 0 (0.0%) | 16 (0.3%) |
| 1 | 1017 (68.7%) | 2420 (82.2%) | 333 (78.9%) | 14 (77.8%) | 24 (70.6%) | 7 (77.8%) | 3815 (77.7%) |
| 2 | 157 (10.6%) | 238 (8.1%) | 33 (7.8%) | 1 (5.6%) | 2 (5.9%) | 0 (0.0%) | 431 (8.9%) |
| 3 | 61 (4.1%) | 122 (4.1%) | 21 (5.0%) | 1 (5.6%) | 2 (5.9%) | 0 (0.0%) | 207 (4.2%) |
| Total | 1481 | 2944 | 422 | 18 | 34 | 9 | 4908 |

**Discussion**

The data we analyze, including the PACT output, inform and reflect a person's path through correctional supervision; this in contrast to previous work by others on a person's path before sentencing and after release. The historic prioritization of security and incapacitation embedded in a tool that shapes the carceral experience is troubling. The PACT was created as a response to the Camp Hill Riot (and the overcrowding that initially sparked it). In lieu of many other pathways, the PACT is part of a comprehensive system to more efficiently manage a population in order to maintain security and control within the prison. Efficiency is prioritized in a shift to a population management approach that is less individualized. The increased reliance on "smart" algorithmic tools serves a reductive function for each individual in the care of the PADOC. Further, the desire for security and control as the prison population boomed highlights the PACT as an algorithm that prioritizes incapacitation and population control – like many algorithms used in corrections.

Our qualitative review of the PACT's historical context invites a "reading for absence,"

as we look at what the PACT collects and analyzes and to what end. It is important to consider what it does not collect and analyze to other ends. Incapacitation is the driver of the collection and weighting of the data. This means much is lost along lines of individuality that could better reveal one's potential as a member of society. What cultural and social assets do incarcerated individuals bring to the prison? What skills do they have that could be developed to ensure future success and reduce recidivism? Of the minimal data of this kind that is collected, none of it appears in the PACT calculation. We can imagine, for example, mental health data might be kept differently and mental health would be treated differently if the algorithmic priority were rehabilitation instead of incapacitation. A tool that prioritized rehabilitation from the start would collect vastly different variables and enact a considerably different future. While there are strong justifications for the importance of security, it must be ceded as the priority and as structuring the PACT at the expense of other priorities. Thus, we find the PACT is not an objective tool, it is a historically and socially constituted tool that reflects subjective decisions and values - specifically security and control. Historicizing the PACT through qualitative analysis helps to invite "reading for absence" (Hesse-Biber, 2012) and gives a window into the limitations of the tool. While we find security to be a primary focus, future research should further examine what is *not* included in carceral algorithms.

The consequences of such absences are best illuminated if the PACT is thought of as both a data point and an algorithm. In our tracing of the PACT, we seek to further advance a study of data and algorithms as iteratively co-productive of one another. By considering the data the PACT informs and the PACT as a future data point, we can better understand how bias gets baked into the varied nodes of the criminal justice system and we can begin to understand feedback loops between the various algorithms deployed in the criminal justice system. Our analysis reveals how PACT perpetuates racial disparity and reduces the humanity of incarcerated people through a data-driven placement process. In examining PACT and the algorithmic process that produces this data point for an incarcerated person, we also can understand the compounding problems in other algorithms (such as parole and sentencing algorithms) that are flawed by extension of relying on the PACT variable.

Our specific attention to the PADOC data about the people it incarcerates reveals the myth of data driven approaches to criminal justice. In practice, the data is replete with issues. The data is not consistently or uniformly collected over time or in discrete periods. The data is also incomplete. We find that over time there are more and more variables being added to the system, but the collection of the data remains inconsistent and error riddled. We find that the PADOC is overconfident in the quality of the data. The actuality of the data reveals that the PACT score is not only reductive, it is also unlikely to generate a reliable calculation. So, we find that the PACT is an algorithm that prioritizes control, but we also find that the data likely undermines the effectiveness of even this embedded goal.

In addition to the unreliability of the data, we reveal biases in the data. In turning to the PACT score as a data point - we find it is replete with racial bias. First in its calculations, we find that Black inmates are more likely to receive higher custody levels. Second, when manual overrides are made, we again see this favoring white inmates. White inmates are more likely to find themselves in a lower custody level and to avoid the need for a manual override.

Manual overrides are frequently used and this is disconcerting in relation to the claims of algorithmic certainty made by smart on crime proponents. On the one hand, overrides could signal that the PACT is not accurate. On the other, overrides could signal that even if the algorithm does reduce bias, human users are given the opportunity to add it back in. In either instance, overrides highlight the human relationality for the algorithm and subsequent constitution of data.

Given all that we find and all the other scholars who have identified bias in the criminal justice system, it is ultimately not a surprise that racial bias in the PACT remains. There is little difference in the PACT scores and calculations over time. Thirty years into a holistic application of algorithm technologies to intake prisoners, the system as a whole remains skewed along racial lines and recidivism rates remain high. Our findings point out that even in the least transparent part of a person's interaction with the PADOC the data is problematic, the decisions are biased and its usefulness is questionable.

A close examination of the PACT reveals the decades long iterations of this practice and the continued stake in both more data and more sophistication as improving the system as such. However, without addressing the issues embedded in mass incarceration, the racially disparate criminal justice system that targets and punishes the poor and people of color, this technology simply reproduces the same, entrenched inequalities through a different, emergent form of supervision. Technological sophistication, the "smart" approach, ultimately works to create a false sense of confidence in the equity and justice of the "data-driven" decisions being made. When one examines the variables that inform prisoner classification, the consistent finding that technological sophistication both obscures and cements bias, is rooted in the limitations of the data. These limitations are rooted in their inherent humanness and the history of their original construction.

Our findings to date further caution against data-driven criminal justice reforms that rely on pre-existing data infrastructures. While the PACT closely informs prisoner rehabilitation plans and parole considerations, it is rooted in population management for prison securitization. Following from O'Neil, we argue that even within the bounds of this system, only some data are examined for specific purposes and objectives that are more consistent with maintaining and justifying the status quo rather than improving or critiquing the system. Continued work is needed to understand the algorithm itself and how it informs parole. So is work that considers data and algorithms that explicitly seek to reduce bias, encourage rehabilitation and reduce recidivism. Such data and algorithms must also interrogate the underlying data infrastructure, given the long history of incomplete data and embedded bias.